\documentclass[runningheads]{llncs}
\usepackage{graphicx}
\usepackage[english]{babel}
\usepackage{gensymb}
\usepackage{caption}
\usepackage{subfig}
\begin{document}
\title{Learned Pre-Processing for Automatic Diabetic Retinopathy Detection on Eye Fundus Images}
\titlerunning{Learned Pre-Processing for DR Detection}
%
\author{Asim Smailagic\inst{1} \and
Anupma Sharan\inst{1} \and
Pedro Costa\inst{2} \and
Adrian Galdran\inst{3} \and
Alex Gaudio\inst{1} \and
    Aur\'{e}lio Campilho\inst{2, 4}}
\authorrunning{Smailagic et al.}
%
\institute{Carnegie Mellon University \and 
INESC TEC, Porto \and 
\'{E}cole de Tecnologie Superieure, Montreal \and 
Faculty of Engineering, University of Porto}
%
\maketitle              
\begin{abstract}
Diabetic Retinopathy is the leading cause of blindness in the working-age population of the world. 
The main aim of this paper is to improve the accuracy of Diabetic Retinopathy detection by implementing a shadow removal and color correction step as a preprocessing stage from eye fundus images.
For this, we rely on recent findings indicating that application of image dehazing on the inverted intensity domain amounts to illumination compensation.
Inspired by this work, we propose a Shadow Removal Layer that allows us to learn the pre-processing function for a particular task.
We show that learning the pre-processing function improves the performance of the network on the Diabetic Retinopathy detection task.

\keywords{Retinal Image Preprocessing \and Diabetic Retinopahty Detection \and Color balancing.}
\end{abstract}
\section{Introduction}

Diabetic Retinopathy (DR) is an eye disease that affects more than $25\%$ of the estimated 425 worldwide diabetic patients \cite{ruta2013prevalence}. 
Consequently, DR is a leading cause of blindness in the working-age population of the world and, therefore, screening all diabetic patients is of paramount importance.
With the growth in the prevalence of diabetes, the burden on ophthalmologists to screen the entire diabetic population also grows.
For these reasons, a system capable of detecting DR is becoming increasingly important.

Screening for DR in the US and UK relies mainly on the right interpretation of a digital retinal image to recognize pathological features. 
Prompt acknowledgement and treatment of this pathology can save sight, and for this reason much research has been devoted in recent years to the design of machine learning pipelines that can help in its correct diagnosis.
Unfortunately, lesions that characterize early stages of this disease are subtle, and when improperly illuminated by a fundus camera in acquisition time they can be confounded with other non-harmful signs of similar appearance. 

A reasonable approach to deal with this problem is to improve the quality of the image by obtaining a shadow free version of the image. 
Although this step can be performed in a manual way \cite{foracchia_luminosity_2005,savelli_illumination_2017,saha_novel_2018}, it may be preferable to learn the preprocessing function with minimal intervention, directly from the data. 
We propose to do so by implementing a U-net architecture \cite{DBLP:journals/corr/RonnebergerFB15}, which is the convolutional neural network architecture of choice for biomedical image segmentation.

\section{Related Work}
The pre-processing of retinal images has been proposed in several papers before.
One of the first proposed techniques for improving the visual appearance of this kind of data was introduced in \cite{foracchia_luminosity_2005}. The authors estimated an illumination field by first removing foreground pixels and then fitting a Gaussian model to the background.
Similarly, the technique proposed in \cite{leahy_illumination_2012} relies on Laplace interpolation and a multiplicative model of illumination to remove its impact.
In \cite{xiong_enhancement_2017}, an image formation model involving scattering and background illumination was proposed and inverted to retrieve well-illuminated images.
A different model, based on cataracts formation, was used in \cite{mitra_enhancement_2018} to reduce blurriness and improve contrast.
Also recently, the authors of \cite{saha_novel_2018} introduce a luminosity correction technique with a focus on avoiding the creation of visual artifacts on regions of the image that were initially well-illuminated.
It is important to stress that all these methods are designed and applied on retinal images in a static manner. This means that any subsequent automatic image understanding task for diagnostic purposes remains isolated from the pre-processing stage.
\begin{figure*}[t]
\centering
\subfloat{\includegraphics[width = 0.35\textwidth]{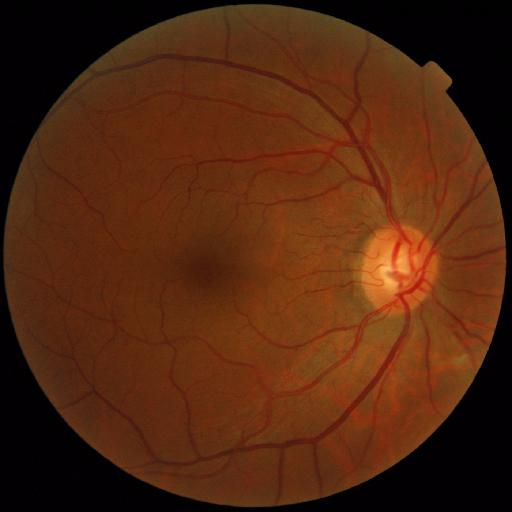}
\label{fig_left_shadow}}
\hfil
\subfloat{\includegraphics[width = 0.35\textwidth]{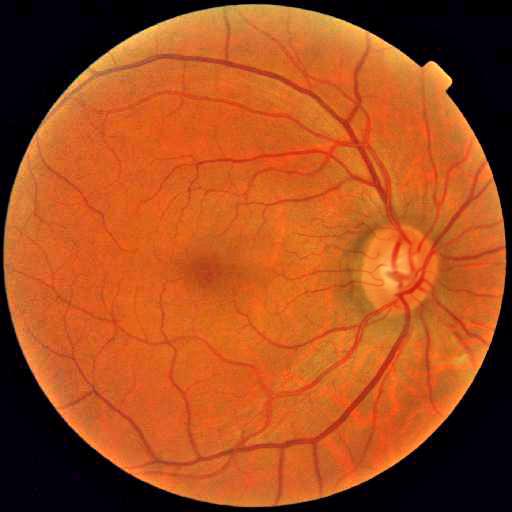}
\label{fig_consensus_left}}
\caption{Example of an eye fundus image. Left: Unprocessed retinal image. Right: Illumination-compensation by shadow removal.}
\label{fig0}
\end{figure*}

In this paper, we follow previous observations from \cite{savelli_illumination_2017,galdran_duality_2018} that fog/haze removal can be interpreted as illumination compensation when applied to inverted intensities on retinal images, as shown in Fig. \ref{fig0}.
The standard model used to describe hazy images is given by the haze imaging equation \cite{Narasimhan2002,Narasimhan00chromaticframework,Fattal:2008:SID:1360612.1360671,inproceedings}:\\
\begin{equation}
    I(x) = J(x)t(x) + A(1 - t(x)).
\end{equation} 
Therefore, haze removal involves estimating the transmission map $t$ (depth map), soft matting for its refinement, estimating the atmospheric light $A$ and recovering the scene radiance $J$.
While we also aim to apply the above model, in contrast with previous techniques our goal in this paper is to automatically learn to estimate these unknowns in such a way that they are optimal for the downstream task of diabetic retinopathy detection, which will be simultaneously solved.

\begin{figure}[t]
  \centering
    \includegraphics[width=0.99\textwidth]{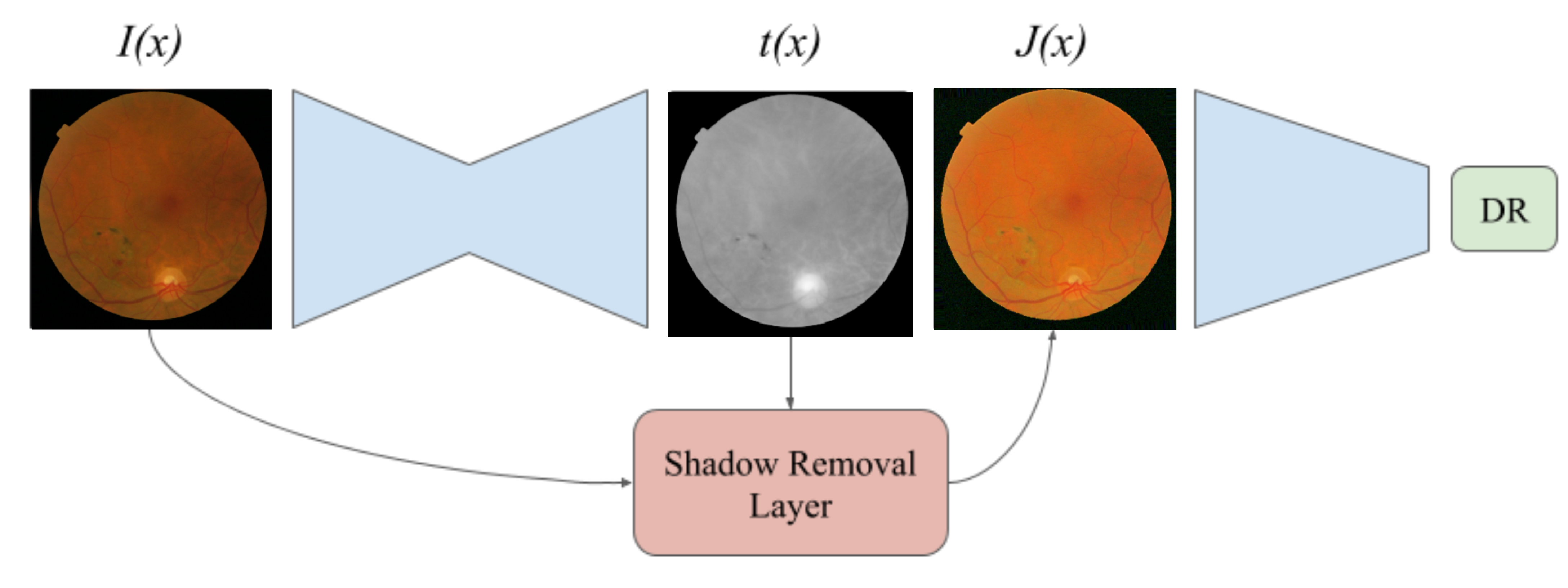}
    \caption{Pipeline of the proposed method. A segmentation CNN is used to estimate the transmission map $t(x)$. Then, the input image $I(x)$ and $t(x)$ are provided to a Shadow Removal Layer that outputs the normalized image $J(x)$. Finally, $J(x)$ is given as input to a classifier CNN that outputs if the image has Diabetic Retinopathy or not. Both CNNs can be trained to minimize the classification error.}
    \label{pipeline}
\end{figure}

\section{Method}

Pre-processing the images to have more consistent illumination and colors across the dataset can help improve the performance of DR detection. 
In this paper, we aim to remove shadows from eye fundus images by dehazing the inverted image \cite{savelli_illumination_2017}.
Dehazing methods require the estimation of the transmission map $t$ using heuristics that may not be optimal.
To overcome this issue, we pose the problem of transmission map estimation as an optimization problem, and propose to
learn the function that maps an eye fundus image to a transmission map $t(x)$ by minimizing a classification error.
This allows us to optimize the transmission map estimation for a particular classification task.

\subsection{Shadow Removal Layer}

In order to accomplish this, we develop a Shadow Removal Layer.
This layer uses an estimated transmission map $t(x)$ and an input image $I(x)$ and outputs a pre-processed image $J(x)$, with shadows removed.
This layer applies the following equation:

\begin{equation}
    J(x) = 1 - (\frac{(1-I(x)) - A}{t(x)} + A).
\label{eq:colorbalance}
\end{equation}

We can assume that $A = 1$ if we white balance the images before applying the illumination estimation function as shown in \cite{savelli_illumination_2017}. The equation then reduces to $I(x) / t(x)$ \textit{i.e.} simply dividing the input image intensities with the transmission map.
In this work, we use a Segmentation Convolutional Neural Network (CNN) to learn the function $t(x)$.

The problem is that we do not have the ground-truth data to train the segmentation model $t(x)$.
To solve this issue, we derive the training signal from a classification CNN that learns to detect DR from $J(x)$.
Therefore, the segmentation CNN learns to output the transmission map that minimizes the classification CNN error.
This is possible as Equation \ref{eq:colorbalance} is differentiable, and the training signal can flow to the segmentation CNN's parameters. 
The entire architecture is shown in Figure \ref{pipeline}.

%

\subsection{Transmission Map Supervision}

For the transmission map estimation model to be able to learn something close to the depth map of the image, we add a term to the loss. 
On top of the classification loss we minimize the mean squared error between $t(x)$ and a reference transmission map $M$. 
This reference transmission map is obtained by computing the depth maps for each image in the dataset manually as per the Dark Channel Prior theory \cite{He:2011:SIH:2068459.2068579} and taking an average over all the depth maps, as shown in Figure \ref{fig:avgt}. 
The objective of the network is hence modified to decrease the difference between the manually computed reference depth map and the learned depth map. The new loss function is:

\begin{figure}[t]
  \centering
    \includegraphics[width=0.4\textwidth]{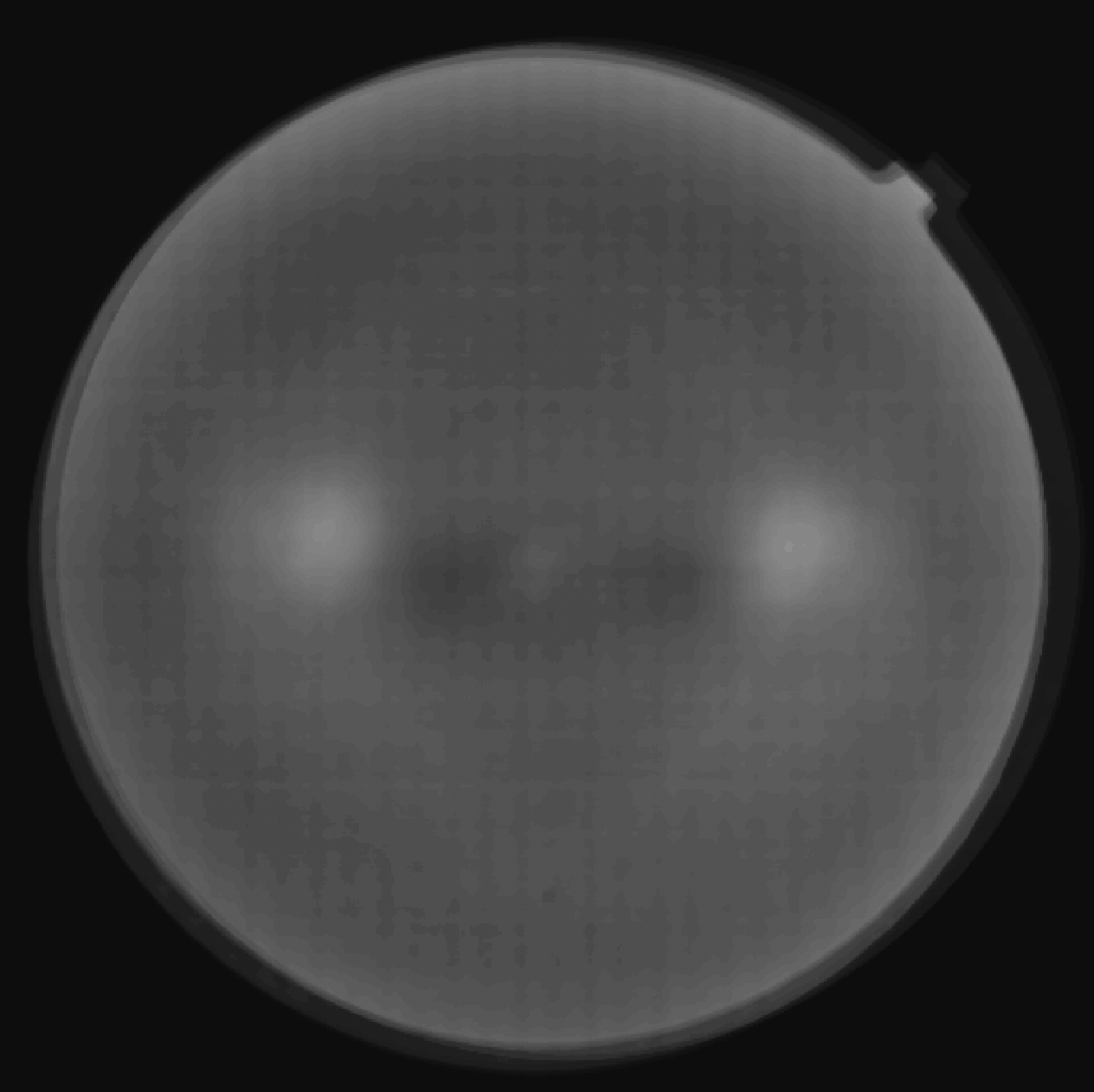}
        \caption{Average depth map computed manually from the entire dataset of eye fundus images, used as additional supervision to the transmission map.}
        \label{fig:avgt}
\end{figure}

\begin{equation}
    J(x, \theta_c, \theta_s) = \mathcal{L}(x, \theta_c, \theta_s) + MSE(x, \theta_s),
\end{equation}
where $\mathcal{L}$ is the classification loss, $\theta_c$ are the classification network's parameters and $\theta_s$ are the segmentation network's parameters. In this paper we used Binary Cross-Entropy as the classification loss $\mathcal{L}$.

\section{Experiments}

\subsection{Implementation Details}

We used a network inspired by U-Net as the segmentation CNN that estimates the transmission map $t(x)$ and a pre-trained Inception v3 network as the classification network. 
The eye fundus images are resized to $512\times512$ and provided to the U-Net. 
The pre-processed images that are given to the Inception v3 network are also $512\times512$.
To accomodate for the larger input image size, we remove the last layer of the Inception v3 network and add a global average pooling layer followed by a Fully-Connected layer with a single output.

The two models are trained using the Adam optimizer with a learning rate of 2x$10^{-4}$. 
The training process consists of 2 phases:
\begin{enumerate}
    \item Fitting: Here, the parameters of the Inception network are frozen and the U-net alone is trained;
    \item Fine-tuning: Here, the layers of the Inception network are made trainable and thus fine tuned along with the U-net parameters
\end{enumerate}
Both fitting and fine-tuning are performed for 100 epochs each with a batch-size of 4. 

\begin{table}[b]
\centering
\begin{tabular}{|l|l|l|}
\hline
Results                                & Test Accuracy \\ \hline
Inception V3                           & 89.50\%       \\ \hline
U-Net+Shadow Removal+Inception v3      & 90.34\%       \\ \hline
\end{tabular}\\
\caption{Our Shadow Removal Layer improves the classification accuracy over the baseline.}
\label{results}
\end{table}

\subsection{Dataset}
The Messidor dataset 
\cite{decenciere_feedback_2014} is a collection of eye fundus images of healthy and unhealthy patients. 
It consists of $1200$ eye fundus color numerical images acquired by 3 ophthalmologic departments. The image sizes are $1440\times960$, $2240\times1488$ or $2304\times1536$ pixels. The retinopathy grade has been provided by medical experts, where a grade of 0 corresponds to healthy and grades 1,2 and 3 correspond to unhealthy.

The training data consists of 949 images, 441 healthy and 508 unhealthy, and the test data consists of 238 images, 106 healthy and 132 unhealthy. The images in both training an test set are distributed equally among the 3 opthalmologic departments.

The images are center cropped and resized. Each image corresponds to 4 images, the original image, a randomly rotated image by an angle in the range of 230$\degree$, and the horizontally flipped version of both.

\subsection{Results}

We trained the classifier on the original dataset for the task of DR detection and obtained $89.50\%$ accuracy.
Our pre-processing method achieves a test accuracy of $90.34\%$ after fine-tuning, as shown in Table \ref{results}, giving an improvement $0.84\%$ in the test set over the baseline. 
Our model converges better than the baseline and also improves the detection.

\begin{figure}[t]
  \centering
        \includegraphics[width=0.4\textwidth]{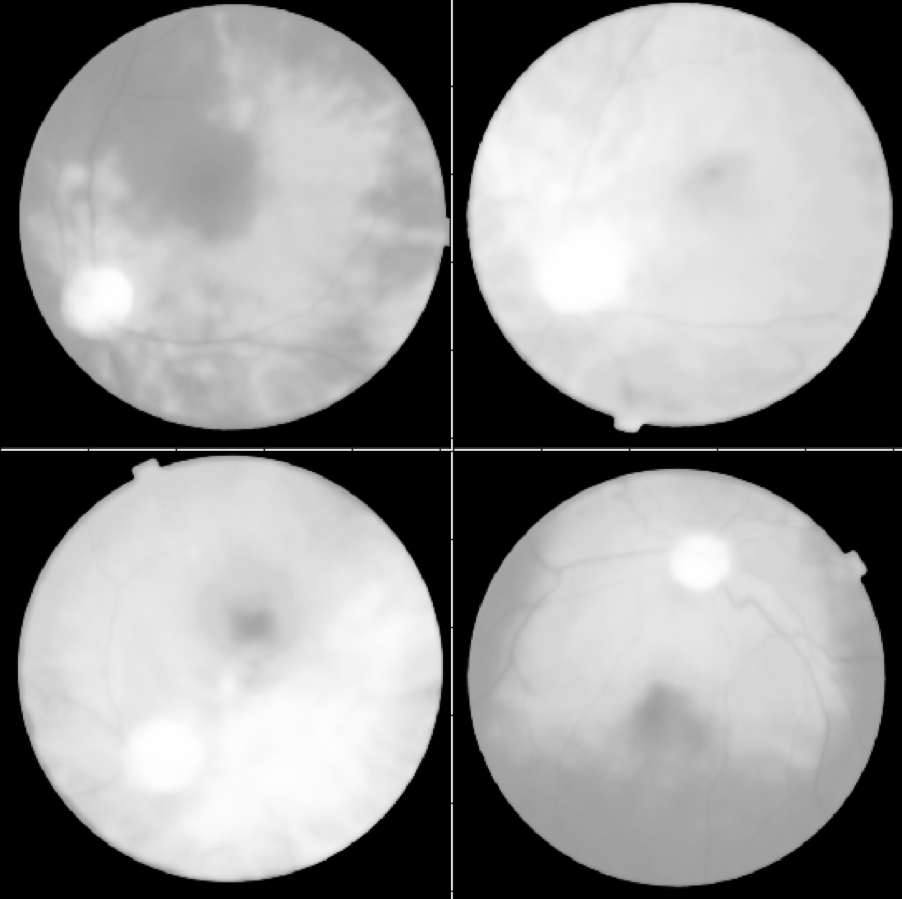}
        \includegraphics[width=0.4\textwidth]{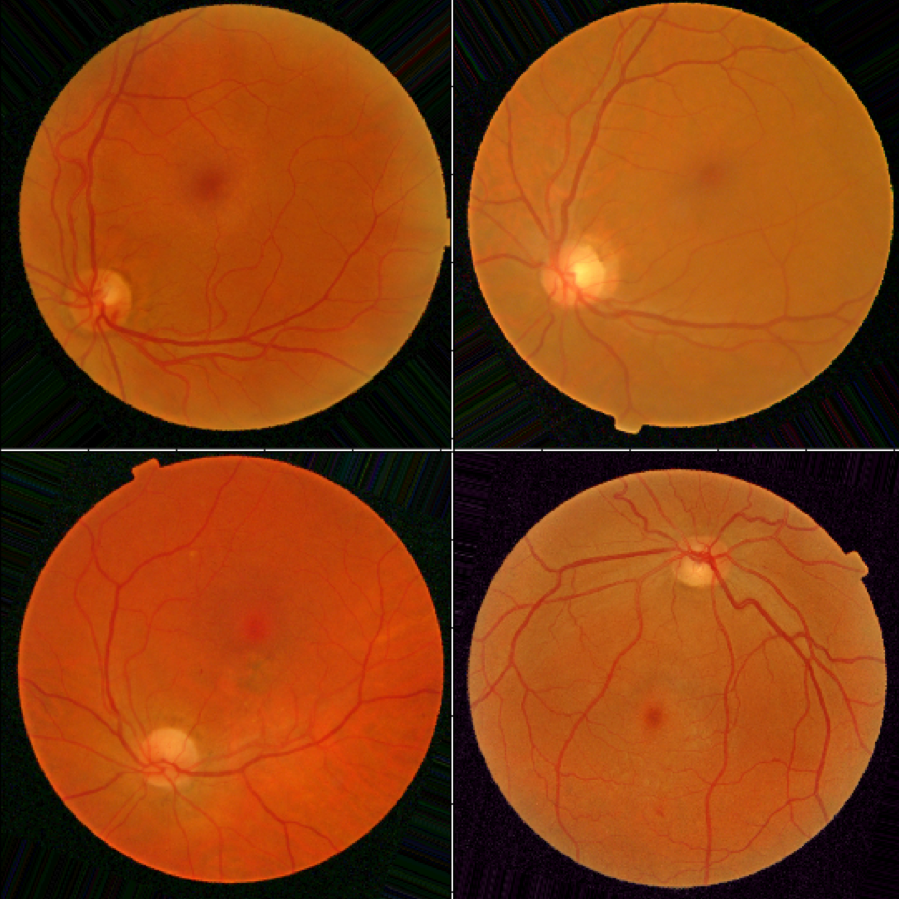}
    
    \caption{Transmission map learned by the U-Net and corresponding output of the Shadow Removal Layer.}
     \label{fig:qualitative}
\end{figure}

Furthermore, we can visually inspect the estimated transmission maps $t(x)$. As shown in Figure \ref{fig:qualitative}, we can verify that the U-Net was able to output valid transmission maps, different from the mean transmission map $M$. 
Moreover, the images produced by the Shadow Removal Layer have similar illumination, indicating that the learned pre-processing step is effectively removing shadows from the eye fundus images.

\section{Conclusion and Future Work}

In this paper we proposed a method to learn how to pre-process eye fundus images for the task of DR detection.
We draw inspiration from haze/shadow removal methods and devise a methodology to train a segmentation CNN to estimate the transmission map of the input eye fundus image.
Then, we apply a Shadow Removal Layer to pre-process the input image and then provide that image to a classifier. 
The entire system can be trained to minimize the classification error.

We show that, by learning to pre-process eye fundus images to a particular task, the performance of DR detection is improved.
As future work, we plan to verify if the learned pre-processing function is useful for other retinal tasks, such as vessel segmentation.

\section*{Acknowledgments}

This work is financed by the ERDF - European Regional Development Fund through the Operational Programme
for Competitiveness and Internationalisation - COMPETE 2020 Programme, by National Funds through the FCT -
Funda\c{c}\~{a}o para a Ci\^{e}ncia e a Tecnologia (Portuguese Foundation for Science and Technology) within project CMUP-ERI/TIC/0028/2014.

\bibliographystyle{unsrt}
\bibliography{ref}

\begin{thebibliography}{10}

\bibitem{ruta2013prevalence}
LM~Ruta, DJ~Magliano, R~Lemesurier, HR~Taylor, PZ~Zimmet, and JE~Shaw.
\newblock Prevalence of diabetic retinopathy in type 2 diabetes in developing
  and developed countries.
\newblock {\em Diabetic medicine}, 30(4):387--398, 2013.

\bibitem{foracchia_luminosity_2005}
Marco Foracchia, Enrico Grisan, and Alfredo Ruggeri.
\newblock Luminosity and contrast normalization in retinal images.
\newblock {\em Medical Image Analysis}, 9(3):179--190, June 2005.

\bibitem{savelli_illumination_2017}
B.~Savelli, A.~Bria, A.~Galdran, C.~Marrocco, M.~Molinara, A.~Campilho, and
  F.~Tortorella.
\newblock Illumination {Correction} by {Dehazing} for {Retinal} {Vessel}
  {Segmentation}.
\newblock pages 219--224, June 2017.

\bibitem{saha_novel_2018}
Sajib Saha, Alexander Fletcher, Di~Xiao, and Yogesan Kanagasingam.
\newblock A novel method for automated correction of non-uniform/poor
  illumination of retinal images without creating false artifacts.
\newblock {\em Journal of Visual Communication and Image Representation},
  51:95--103, February 2018.

\bibitem{DBLP:journals/corr/RonnebergerFB15}
Olaf Ronneberger, Philipp Fischer, and Thomas Brox.
\newblock U-net: Convolutional networks for biomedical image segmentation.
\newblock {\em CoRR}, abs/1505.04597, 2015.

\bibitem{leahy_illumination_2012}
Conor Leahy, Andrew O’Brien, and Chris Dainty.
\newblock Illumination correction of retinal images using {Laplace}
  interpolation.
\newblock {\em Applied Optics}, 51(35):8383--8389, December 2012.

\bibitem{xiong_enhancement_2017}
Li~Xiong, Huiqi Li, and Liang Xu.
\newblock An enhancement method for color retinal images based on image
  formation model.
\newblock {\em Computer Methods and Programs in Biomedicine}, 143:137--150, May
  2017.

\bibitem{mitra_enhancement_2018}
Anirban Mitra, Sudipta Roy, Somais Roy, and Sanjit~Kumar Setua.
\newblock Enhancement and restoration of non-uniform illuminated {Fundus}
  {Image} of {Retina} obtained through thin layer of cataract.
\newblock {\em Computer Methods and Programs in Biomedicine}, 156:169--178,
  March 2018.

\bibitem{galdran_duality_2018}
A.~Galdran, A.~Bria, A.~Alvarez-Gila, J.~Vazquez-Corral, and M.~Bertalmío.
\newblock On the {Duality} {Between} {Retinex} and {Image} {Dehazing}.
\newblock In {\em 2018 {IEEE}/{CVF} {Conference} on {Computer} {Vision} and
  {Pattern} {Recognition}}, pages 8212--8221, June 2018.

\bibitem{Narasimhan2002}
Srinivasa~G. Narasimhan and Shree~K. Nayar.
\newblock Vision and the atmosphere.
\newblock {\em International Journal of Computer Vision}, 48(3):233--254, Jul
  2002.

\bibitem{Narasimhan00chromaticframework}
Srinivasa~G. Narasimhan and Shree~K. Nayar.
\newblock Chromatic framework for vision in bad weather.
\newblock In {\em In Proceedings of the IEEE Conference on Computer Vision and
  Pattern Recognition}, 2000.

\bibitem{Fattal:2008:SID:1360612.1360671}
Raanan Fattal.
\newblock Single image dehazing.
\newblock {\em ACM Trans. Graph.}, 27(3):72:1--72:9, August 2008.

\bibitem{inproceedings}
Robby Tan.
\newblock Visibility in bad weather from a single image.
\newblock 06 2008.

\bibitem{He:2011:SIH:2068459.2068579}
Kaiming He, Jian Sun, and Xiaoou Tang.
\newblock Single image haze removal using dark channel prior.
\newblock {\em IEEE Trans. Pattern Anal. Mach. Intell.}, 33(12):2341--2353,
  December 2011.

\bibitem{decenciere_feedback_2014}
Etienne Decencière, Xiwei Zhang, Guy Cazuguel, Bruno Lay, Béatrice Cochener,
  Caroline Trone, Philippe Gain, Richard Ordonez, Pascale Massin, Ali Erginay,
  Béatrice Charton, and Jean-Claude Klein.
\newblock Feedback on a publicly distributed database: the messidor database.
\newblock {\em Image Analysis \& Stereology}, 33(3):231--234, August 2014.

\end{thebibliography}
\end{document}